\documentclass[reprint, aps, prl, superscriptaddress,nobibnotes]{revtex4-1}

\usepackage{graphicx}
\usepackage{dcolumn}
\usepackage{amsmath}
\usepackage{textcomp}
\usepackage{bm}
\usepackage{braket}
\usepackage[left]{lineno}
\usepackage{color,soul}

\newcommand{\highlite}[1]{\textcolor{black}{#1}}

\begin{document}

\title{Density-controlled quantum Hall ferromagnetic transition in a two-dimensional hole system}

\author{T. M. Lu}
\email{tlu@sandia.gov}
\affiliation{Sandia National Laboratories, Albuquerque, New Mexico 87185, USA}
\author{L. A. Tracy}
\affiliation{Sandia National Laboratories, Albuquerque, New Mexico 87185, USA}
\author{D. Laroche}
\affiliation{Sandia National Laboratories, Albuquerque, New Mexico 87185, USA}
\author{S.-H. Huang}
\affiliation{Department of Electrical Engineering and Graduate Institute of Electronic Engineering, National Taiwan University, Taipei 10617, Taiwan, R.O.C.}
\affiliation{National Nano Device Laboratories, Hsinchu 30077, Taiwan, R.O.C.}
\author{Y. Chuang}
\affiliation{Department of Electrical Engineering and Graduate Institute of Electronic Engineering, National Taiwan University, Taipei 10617, Taiwan, R.O.C.}
\affiliation{National Nano Device Laboratories, Hsinchu 30077, Taiwan, R.O.C.}
\author{Y.-H. Su}
\affiliation{Department of Electrical Engineering and Graduate Institute of Electronic Engineering, National Taiwan University, Taipei 10617, Taiwan, R.O.C.}
\affiliation{National Nano Device Laboratories, Hsinchu 30077, Taiwan, R.O.C.}
\author{J.-Y. Li}
\affiliation{Department of Electrical Engineering and Graduate Institute of Electronic Engineering, National Taiwan University, Taipei 10617, Taiwan, R.O.C.}
\affiliation{National Nano Device Laboratories, Hsinchu 30077, Taiwan, R.O.C.}
\author{C. W. Liu}
\affiliation{Department of Electrical Engineering and Graduate Institute of Electronic Engineering, National Taiwan University, Taipei 10617, Taiwan, R.O.C.}
\affiliation{National Nano Device Laboratories, Hsinchu 30077, Taiwan, R.O.C.}

\date{\today}

\begin{abstract}

\highlite{Quantum Hall ferromagnetic transitions} are typically achieved by increasing the Zeeman energy through in-situ sample rotation, \highlite{ while transitions in systems with pseudo-spin indices can be induced by gate control.}  We report here a \highlite{gate}-controlled quantum Hall ferromagnetic transition \highlite{between two real spin states} in a conventional two-dimensional system without any in-plane magnetic field.  We show that the ratio of the Zeeman splitting to the cyclotron gap in a Ge two-dimensional hole system increases with decreasing density owing to inter-carrier interactions.  Below a critical density of $\sim2.4\times 10^{10}$ cm$^{-2}$, this ratio grows greater than $1$, resulting in a ferromagnetic ground state at filling factor $\nu=2$.  \highlite{At the critical density, a resistance peak due to the formation of microscopic domains of opposite spin orientations is observed.}  Such gate-controlled spin-polarizations in the quantum Hall regime opens the door to realizing Majorana modes using two-dimensional systems in conventional, low-spin-orbit-coupling semiconductors.  

\end{abstract}

\maketitle
In a perpendicular magnetic field ($B_{p}$), two-dimensional (2D) electrons/holes execute cyclotron motion, and their energy spectrum consists of a ladder of discrete Landau levels\cite{Ando1982}.
The spacing between two neighboring Landau levels is given by the cyclotron gap $E_{c}={\hbar eB_{p}}/{m^{*}m_{0}}$, where $\hbar$ is the reduced Planck's constant, $e$ is the elementary charge, and $m^{*}$ is the in-plane effective mass in units of $m_{0}$, the free-electron mass.  
Each Landau level further splits into two spin-polarized levels due to the Zeeman effect with a Hamiltonian $H_S=\frac{1}{2}\mu_{B}\boldsymbol{\sigma gB}$, where $\mu_{B}=e\hbar/2m_{0}$ is the Bohr magneton, $\boldsymbol{\sigma}$ is the Pauli operator, $\boldsymbol{g}$ is the Land\'{e} $g$ tensor, and $\boldsymbol{B}$ is the total magnetic field, which includes $B_{p}$ and potentially an in-plane magnetic field ($B_{ip}$).  
For most semiconductors, the $g$ tensor can be approximately reduced to two components, $g_{p}$ and $g_{ip}$, for the perpendicular magnetic field and the in-plane magnetic field respectively.  
In this case, the energy splitting due to the Zeeman effect is $E_{z}=((g_{p}\mu_{B}B_{p})^2+(g_{ip}\mu_{B}B_{ip})^2)^\frac{1}{2}$, and $E_{z}=g_{p}\mu_{B}B_{p}$ in the absence of an in-plane magnetic field.

Each spin-polarized Landau level has a degeneracy of $eB_{p}/h$ and can be labeled by its spin orientation, $\uparrow$ or $\downarrow$, and an integer $n$ for its Landau level index.  
Here, we use $\downarrow$ to label the lower-energy level.  
The ordering of and the energy gaps between these spin-polarized Landau levels depend on the ratio of the Zeeman splitting to the cyclotron gap, $r\equiv E_{z}/E_{c}$\cite{Nicholas1983}.
In the case of $r<1/2$, $\ket{n\uparrow}$ and $\ket{n\downarrow}$ are close in energy while other Landau levels are far away.  
A schematic drawing of such a spectrum is shown in the top panel of Fig.~1a.  
The middle panel of Fig.~1a shows the alignment of the Landau levels at $r\sim1/2$.  
In this case all the Landau levels are equally spaced.  
In the case of $1/2<r<1$, the energy gap between between $\ket{n\uparrow}$ and $\ket{n+1\downarrow}$ becomes smaller than the the gap between $\ket{n\uparrow}$ and $\ket{n\downarrow}$, as is shown in the bottom panel of Fig.~1a.  
The uncommon but achievable situation of $r>1$ corresponds to reverse ordering of $\ket{n\uparrow}$ and $\ket{n+1\downarrow}$ in energy.  
At $r\sim1$, $\ket{n\uparrow}$ and $\ket{n+1\downarrow}$ are roughly degenerate, and this regime is the main focus of our work.

When $B_{p}$ is sufficiently high such that the energy gaps between the Landau levels are larger than the disorder broadening of the levels, one can observe Shubnikov-de Haas oscillations and the integer quantum Hall effect through magneto-resistance measurements\cite{Ando1982, Klitzing1980}.
For a 2D system with a carrier density $p$, the number of filled Landau levels (filling factor $\nu$) is $hp/eB_{p}$.  
When $\nu$ is tuned to an integer by varying $B_{p}$ and/or $p$, a minimum in the longitudinal magneto-resistance ($R_{xx}$) and quantization of the transverse magneto-resistance ($R_{xy}$) occur.  
A large energy gap between the topmost filled Landau level and the lowest empty Landau level leads to a deep resistance minimum in $R_{xx}$, eventually developing to zero resistance.  
Without an in-plane magnetic field, $r$  is simply $m^{*}g_{p}/2$.    
Most n-type semiconductors have $g_{p}\sim g_{ip}=g^{*}$ and a product of $m^{*}$ and $g^{*}$ smaller than $1$ \cite{Winkler2003}, yielding larger energy gaps when $\nu$ is even rather than odd.  
Typical $R_{xx}$ traces in these materials thus first show strong minima at even filling factors.
Weaker minima at odd filling factors only appear after the disorder broadening no longer masks the Zeeman splitting at sufficiently high magnetic fields.  
\highlite{By applying an in-plane magnetic field, which enhances $E_{z}$ while keeping $E_{c}$ constant, $r$ can be artificially enhanced in these materials.}
When $r$ is tuned to a value greater than $1/2$, the $R_{xx}$ minima at odd filling factors become stronger than when $\nu$ is even.  
In fact, one can even achieve the $r=1,2,3,...$ conditions and observe the sequential closing and re-opening of the energy gap at a fixed $\nu$ as $r$ crosses integer values\cite{Fang1968,Nicholas1983,Koch1993,Papadakis1999,Lai2006a}.
Another way of achieving the crossing condition $r\sim1$ is to change $m^{*}g_{p}$ through many-body effects.
While $m^{*}$ and $g_{p}$ are mostly material parameters set by the band structure, they can be significantly enhanced by interactions, which are controlled by the carrier density\cite{Fang1968,Smith1972,Ando1974,Englert1982}.
\highlite{AlAs\cite{Vakili2004, Shkolnikov2004, Shayegan2006} and Si\cite{Shashkin2001} electrons have been shown to have strongly enhanced $m^{*}g_{p}$, which can be greater than $1$.}

When $E_{F}$ lies between two nearly degenerate Landau levels, electron-electron interactions can lead to interesting quantum phases and/or phase transitions not captured by the simple single-particle picture described above\cite{Jungwirth1998,Piazza1999,DePoortere2000,Spielman2001,Muraki2001,Vakili2006}.
The nature of these phenomena depends on the spatial wavefunctions, real spin indices, and pseudo-spin indices of the two crossing Landau levels, and even the presence of other Landau levels nearby.
For single-subband single-valley 2D systems, the two crossing levels typically have different Landau level indices and opposite real spin indices.
The simplest case, the case of interest in this work, is when the two crossing levels are $\ket{n\uparrow}$ and $\ket{n+1\downarrow}$, which exhibit a ferromagnetic instability near the crossing point\cite{Giuliani1985,Yarlagadda1991,Jungwirth2000}.
While promoting carriers from $\ket{n\uparrow}$ to $\ket{n+1\downarrow}$ costs energy, the 2D system lowers its total energy in this configuration because the gain in exchange energy between $\ket{n+1\downarrow}$ and $\ket{n\downarrow}$ offsets the energy penalty\cite{Giuliani1985}.
With this spin-polarized ground state, the system can be described as an Ising quantum Hall ferromagnetic system\cite{Jungwirth1998}.
Near the crossing point and at finite temperatures, the system breaks up into magnetic domains\cite{Jungwirth2001}.
Electrons moving along the domain wall loops cause dissipation, which is characterized by a sharp resistance peak in $R_{xx}$.
Experimentally, such resistance peaks near the crossing of two Landau levels have been observed in 2D electron systems in several semiconductors, including GaAs\cite{Koch1993}, AlAs\cite{DePoortere2000,DePoortere2003}, InSb\cite{Chokomakoua2004}, Mn-doped CdTe\cite{Jaroszynski2002} and Si\cite{Lai2006b,Toyama2008}, by applying an in-plane magnetic field, typically through in-situ sample rotation.
Recently, a gate-controlled quantum Hall ferromagnetic transition has been reported in Mn-doped CdTe by manipulating the spin-orbit interaction\cite{Kazakov2016}.
\highlite{We also note that gate-controlled quantum Hall ferromagnetic transitions have been observed in systems with pseudo-spins\cite{Muraki2001,Vakili2006}.}
\highlite{In these cases, the electric-subband\cite{Muraki2001} and valley\cite{Vakili2006} indices play the role of spins, and the physical descriptions are similar to the real spin case.}
Recently, we have reported the fabrication and the magneto-transport properties of capacitively induced 2D holes in a Ge quantum well\cite{Laroche2016}.
Utilizing the large band value of $m^{*}g_{p}$ in Ge\cite{Winkler2003}, we report a \highlite{density}-controlled quantum Hall ferromagnetic transition at $\nu=2$ \highlite{between two states with different real spin configurations due to further enhancement of $m^{*}g_{p}$ by interactions.}

\section{Results}
\subsection{Enhancement of $r$ at low densities}

The starting materials for this work are Ge quantum wells sandwiched by SiGe barriers.  
Details of the device fabrication can be found in the Methods section as well as in our previous work\cite{Laroche2016}.
The fabricated heterostructure field-effect transistors allow us to capacitively induce 2D holes in the Ge quantum well with a density as low as $\sim1\times 10^{10}$ cm$^{-2}$ and as high as $\sim2\times 10^{11}$ cm$^{-2}$. 
Magneto-transport measurements were made at temperature $T\sim0.3$ K to obtain the density dependence of $R_{xx}$ and $R_{xy}$ of the induced 2D holes (see Methods).
In Fig.~1b, we show $R_{xx}$ as a function of $B_{p}$ at three densities ($1.91\times 10^{11}$ cm$^{-2}$, $1.57\times 10^{11}$ cm$^{-2}$ and $1.03\times 10^{11}$ cm$^{-2}$ from top to bottom).
At the highest density (top panel), Shubnikov-de Haas oscillations first appear at even filling factors with increasing $B_{p}$.
Resistance minima at odd filling factors appear later and are shallower than at neighboring even filling factors.
At the intermediate density (middle panel), the strength of the oscillations increases monotonically with increasing $B_{p}$, regardless of $\nu$ being even or odd.
At the lowest density (bottom panel), the odd states appear first while the even states are only resolved at larger $B_{p}$ with weaker resistance minima.
Since even states are stronger at $r<1/2$ and odd states are stronger at $r>1/2$, as shown in Fig.~1a, it is apparent that a crossover from $r<1/2$ to $r>1/2$ occurs near the intermediate density $p=1.57\times 10^{11}$ cm$^{-2}$.

\subsection{Density-controlled quantum Hall ferromagnetic transition}

A color plot of $R_{xx}$ as a function of $B_{p}$ and $p$ is shown in the left panel of Fig.~2a, with prominent integer quantum Hall states labeled in yellow.
The most salient feature of the data is a resistance peak at $B_{p}\sim0.5$ T around $p\sim2.4\times 10^{10}$ cm$^{-2}$ in the $\nu=2$ minimum.  
A zoom-in view of this region is shown in the right panel of Fig.~2a.
Figure 2b shows five line cuts along the $B_{p}$ axis near $p\sim2.4\times 10^{10}$ cm$^{-2}$, with the densities labeled on the right.
It can be seen that the $\nu=2$ minimum persists in this range of densities and moves from low to high magnetic field with increasing $p$, as expected.
The anomalous peak appears to be additive to the $\nu=2$ minimum and has a weak $p$ dependence.
We believe that this peak is evidence of microscopic magnetic domains formation arising from a quantum Hall ferromagnetic transition at $\nu=2$ that is induced purely by enhanced interactions.
In Fig.~2c, we plot the peak position ($B_{peak}$) as a function of $p$ in the top panel and convert $B_{peak}$ to a filling factor $\nu_{peak}$, shown in the lower panel.
It is interesting to note that $\nu_{peak}$ is greater than $2$ at $p>2.4\times 10^{10}$ cm$^{-2}$ and is smaller than $2$ at $p<2.4\times 10^{10}$ cm$^{-2}$.

\subsection{Temperature dependence}

The temperature dependence of the ferromagnetic transition peak at $p\sim2.4\times 10^{10}$ cm$^{-2}$ is shown in Fig.~3a.  
To focus on the anomalous peak structure, we subtract a linear background from $R_{xx}$ for each temperature, as shown in the inset of Fig.~3a.
The temperature dependences of the peak height $\Delta R_{peak}$ and the peak width $\Delta B_{p}$ after the removal of the background are displayed in Fig.~3b.
Within our accessible temperature range, $\Delta R_{peak}$ decreases and $\Delta B_{p}$ increases monotonically with increasing $T$.
By $T\sim0.8$ K the peak is barely observable.

\subsection{Magnetic-field sweep rate dependence}

Magnetic domains and their motions are non-equilibrium phenomena, which could manifest themselves in a time dependence of $R_{xx}$ and/or hystereteic behavior.
No hysteresis with respect to magnetic field sweeping direction has been observed down to $T\sim0.3$ K.  
Nevertheless, $R_{xx}$ near the transition depends on the magnetic-field sweep rate $dB_{p}/dt$.
In Fig.~3c, we show $R_{xx}$ at five different magnetic-field sweep rates.
It is apparent that the $\nu=2$ minimum rises with increasing $dB_{p}/dt$, while the peak resistance and the neighboring $\nu=3$ state are insensitive to the sweep rate.
A rising $\nu=2$ minimum and fixed peak resistance translate to decreasing $\Delta R_{peak}$ with increasing $dB_{p}/dt$, which is shown in Fig.~3d.

\section{Discussion}

Enhancement of $m^{*}$ or $g_{p}$ due to interactions has been observed since the early days of 2D electrons in Si metal-oxide-semiconductor field-effect transistors\cite{Fang1968,Smith1972,Ando1974}, and has also been observed in other 2D systems\cite{Englert1982,Vakili2004,Shayegan2006,Tsukazaki2008,Nedniyom2009,Melnikov2014}.
For Ge 2D holes, the band value of $g_{p}$ is mainly determined by the Luttinger parameters \cite{Luttinger1956} and can be as high as $\sim$20\cite{Winkler2003}.  
\highlite{The smooth crossover from $r<1/2$ to $r>1/2$ as $p$ decreases is consistent with observations in other material systems.}
As $p$ keeps decreasing, it is natural to expect $r$ to keep increasing with stronger interactions and to eventually become equal to and then greater than 1.
When this happens, the ordering of the $\ket{0\uparrow}$ and $\ket{1\downarrow}$ states reverses and an Ising quantum Hall ferromagnetic transition occurs.  
The resistance peak inside the $\nu=2$ minimum shown in Fig.~2a is the evidence of such a density-controlled quantum Hall ferromagnetic transition in our system.
Staying along the $\nu=2$ line in the $p$-$B_{p}$ space, the system is in a spin-unpolarized ground state at higher $p$, and hence higher $B_{p}$, as the two filled Landau levels, $\ket{0\downarrow}$ and $\ket{0\uparrow}$, have opposite spins.  
As $B_{p}$ and $p$ decrease, the system transitions to a ferromagnetic ground state, as the two filled Landau levels, $\ket{0\downarrow}$ and $\ket{1\downarrow}$, have parallel spins.
This interesting, counterintuitive behavior can be understood as a result of the exchange interaction, which tends to align the spin of carriers along the same direction.  
This exchange energy can be estimated by comparing the experimentally obtained energy spectrum with that of non-interacting 2D holes.  
Cyclotron resonance yields a band mass of $0.091$ for high-mobility Ge 2D holes\cite{Failla2016}, while $m^{*}$ obtained from mobility analysis is $0.07$ \cite{Dobbie2012}.
Experimentally measured values of $g_{p}$ of high-mobility Ge 2D holes range from $2.8$ to $7.0$ \cite{Failla2016}, smaller than the estimate of $\sim20$ from bulk Luttinger parameters.
These experimentally measured values of $m^{*}$ and $g_{p}$ lead to a gap ranging from $\sim5$ K to $\sim9$ K at $\nu=2$ between $\ket{0\uparrow}$ and $\ket{1\downarrow}$ at $B_{p}=0.5$ T, assuming a normal ordering of Landau levels.
At $p\sim2.4\times 10^{10}$ cm$^{-2}$, the system at $\nu=2$ lowers its energy through exchange interaction by this gap amount per particle, and a transition to a spin-polarized ground state occurs.
The exchange energy term arises from the Coulomb interaction and is expected to be a fraction of the Coulomb energy ($e^2/4\pi\epsilon l_B$, where $\epsilon$ is the dielectric constant of Ge and $l_B$ is the magnetic length $(\hbar/eB_{p})^{1/2}$), the relevant energy scale.
At $B_{p}=0.5$ T the Coulomb energy is $\sim30$ K and is indeed consistent with the above statement.

The configurations of the four lowest spin-polarized Landau levels before and after the transition are shown in Fig.~2d.  
On the low-density side, the lowest empty Landau level shares the same Landau level index with the lowest filled Landau level, but has a different spin index.
On the high-density side, the lowest empty Landau level shares the same spin index with the lowest filled Landau level, but has a different Landau level index.
The density dependence of $\nu_{peak}$ thus exhibits an asymmetry between the Landau level indices and the spin indices.
The degree of asymmetry can be quantified by the slope of $\nu_{peak}$ vs.~$p$ near the transition, which is $\sim0.8/10^{10}$ cm$^{-2}$.
This asymmetry can be attributed to the fact that the exchange energy depends on the actual spatial wavefunctions\cite{Giuliani1985,Jungwirth2000}, which in turn depend on the Landau level indices.
For $\nu_{peak}>2$, upon adding particles to the lowest empty Landau level, the system lowers its energy through exchange interaction with the lowest filled Landau level $\ket{0\downarrow}$.
In the single-particle picture, this lowering in energy is equivalent to $\ket{1\downarrow}$ moving closer to $\ket{0\uparrow}$.
Eventually the states $\ket{1\downarrow}$ and $\ket{0\uparrow}$ cross when enough particles are added, and the system breaks up into magnetic domains, resulting in a resistance peak.
A resistance peak is not observed upon adding more particles on the low-density side, since the system does not gain additional energy through exchange interaction with the filled levels.  
On the other hand, by removing particles from the highest filled Landau level, part of the originally gained exchange energy which stabilizes the ferromagnetic ground state is lost.
The 2D system reverts back to the paramagnetic ground state when too many particles are removed.

Previous studies on quantum Hall ferromagnetic transitions typically show a resistance peak that has a maximum height at an intermediate temperature, which is defined as the Curie temperature of the quantum Hall ferromagnets\cite{DePoortere2000,Jaroszynski2002,DePoortere2003}.
The monotonically decreasing peak height and broadening peak width with increasing temperature shown in Fig.~3b indicate that the Curie temperature in 2D holes in Ge is below $\sim$0.3 K.  
This is not too surprising considering that the energy scales in our system are smaller than in other studies.  
Indeed, the relevant density in our system is $p\sim2.4\times 10^{10}$ cm$^{-2}$, which is approximately ten times smaller than the densities used in previous experiments\cite{DePoortere2000,Jaroszynski2002,DePoortere2003}.
The Fermi energy and Coulomb energy are correspondingly much smaller.
It remains to be seen whether a non-monotonic temperature dependence can be observed at dilution-refrigerator temperatures.
Nevertheless, evidence of magnetic domains and their non-equilibrium nature is revealed by the sweep rate dependence of the resistance peak, shown in Fig.~3c and 3d. 
At higher magnetic-field sweep rates, the background resistance near the peak is more resistive.
This rise in the background resistance at a higher sweep rate can be qualitatively understood in the following way.
At the resistance peak, the system is characterized by magnetic domains with opposite spins which have quickly relaxed to a stable or metastable configuration.
The dynamics at the peak is too fast to be captured by our variable-sweep-rate measurements.
Away from the peak, the system may be characterized by numerous isolated puddles of one spin embedded in a matrix with the opposite spin.
Surface tension drives the system toward a lower energy state by minimizing the number of these microscopic domains by domain wall motion and domain merging.
This two-dimensional ripening dynamics can be slow, especially when the density of puddles is low and distance between them is large, and manifests itself in the sweep rate dependence.
At a faster sweep rate, there are more microscopic domains and more domain walls, resulting in a more resistive background.

A gate-controlled quantum Hall ferromagnetic transition enables the formation of a helical domain wall at the boundary between two oppositely spin-polarized 2D systems controlled by two gates\cite{Kazakov2016}.  
The helical domain walls, when coupled to an s-wave superconductor, could allow conventional semiconductors with weak spin-orbit coupling to support Majorana bound states\cite{Clarke2013}.
\highlite{We note that here the real spins are required for induced superconductivity.}
However, entering the quantum Hall regime and observing a quantum Hall ferromagnetic transition require a magnetic field sufficiently high to resolve the Landau levels.  
The requirement of a strong magnetic field is typically in conflict with superconductivity.
The quantum Hall ferromagnetic transition observed in our system occurs at a small magnetic field, $B_{p}=$ 0.5 T.
Our results thus \highlite{show that it may be possible to realize Majorana modes with gate control in conventional semiconductors with weak spin-orbit coupling.}

In summary, we present a systematic evolution from strong even states to strong odd states as $m^{*}g_{p}$ is enhanced by interaction in Ge 2D holes.
Eventually, this interaction enhancement is so strong that a density-controlled quantum Hall ferromagnetic transition is observed, evidenced by a sharp resistance peak in the $\nu=2$ minimum.
While no hysteresis and non-monotonic temperature dependence are observed within our accessible temperature range, the time dependence of $R_{xx}$ near the resistance peak indicates motion of the magnetic domains.

\section{Methods}
\subsection{Material growth and sample fabrication.} 

Two wafers were used in this study, allowing for a wider accessible density range.
The wafers were grown in a rapid-thermal chemical-vapor-deposition system.  
For each wafer, a thin Ge layer was grown on a Si (100) substrate, followed by a reverse graded buffer layer and a $\sim$3-$\mu$m-thick relaxed buffer layer.  
The two wafers had a Si concentration of 21$\%$ and 27$\%$, respectively.  
The wafer with the higher Si concentration has a larger saturation density because of a higher tunnel barrier between the quantum well and the oxide/semiconductor interface\cite{Laroche2016}.
A $\sim$14-nm-thick strained Ge quantum well and a SiGe cap layer were grown after the relaxed SiGe buffer layer.  
The SiGe cap layers were 484 nm and 70 nm thick for the low and high Si concentration wafers, respectively.  
The thicknesses of the SiGe cap layers were obtained by cross-sectional transmission electron microscopy and secondary ion mass spectroscopy.

We used a process flow similar to that reported in Ref.~\onlinecite{Laroche2016} to fabricate heterostructure field-effect transistors.  
Ohmic contacts were made by annealing 250-nm-thick Al pads deposited directly on the epi-layers at 490$^{\circ}$C in a forming gas environment. 
An insulating Al$_2$O$_3$ layer was then deposited by atomic layer deposition, followed by metal deposition for the gate layer.  
To electrically contact the ohmic contacts, vias were etched and wire bonding pads were deposited. 

\subsection{Measurement setup.}

Magneto-transport measurements were performed in three different $^3$He cryostats, all with a base temperature of $\sim$ 0.3 K.  
The carrier density was tuned by gate voltage without any illumination.
Low-frequency lock-in measurements with an excitation current of 5-10 nA were performed to measure the longitudinal and the Hall resistances.  
The carrier density used in the figures was obtained from the Hall resistance.
The 2D color plot was obtained by sweeping the gate voltage and incrementing the magnetic field after each voltage scan.

\section{Acknowledgements}
We thank A. Suslov, H. Baek, G. Jones, and T. Murphy for their assistance in experiments.  This work has been supported by the Division of Materials Sciences and Engineering, Office of Basic Energy Sciences, U.S. Department of Energy (DOE).  This work was performed, in part, at the Center for Integrated Nanotechnologies, a U.S. DOE, Office of Basic Energy Sciences, user facility.  Sandia National Laboratories is a multi-mission laboratory managed and operated by National Technology and Engineering Solutions of Sandia, LLC., a wholly owned subsidiary of Honeywell International, Inc., for the U.S. Department of Energy's National Nuclear Security Administration under contract DE-NA-0003525.  A portion of this work was performed at the National High Magnetic Field Laboratory, which is supported by National Science Foundation Cooperative Agreement No. DMR-1157490 and the State of Florida.  The Ge/SiGe heterostructures were prepared by NTU and supported by the Ministry of Science and Technology (103-2112-M-002-002-MY3 and 105-2622-8-002-001).

\section{Author contributions}
S.H.H., Y.C., J.Y.L., and C.W.L. performed the SiGe epitaxial growth and characterization.  
T.M.L, D.L. and Y.H.S. fabricated the heterostructure field-effect transistor devices.  
T.M.L, L.A.T., D.L., and Y.H.S. performed the low-temperature experiments and data analysis.  
All authors contributed to the writing of the manuscript.

\section{Competing financial interests}
The authors declare no competing financial interests.

\newpage

\begin{figure}[h]
\resizebox{6 in}{!}{\includegraphics{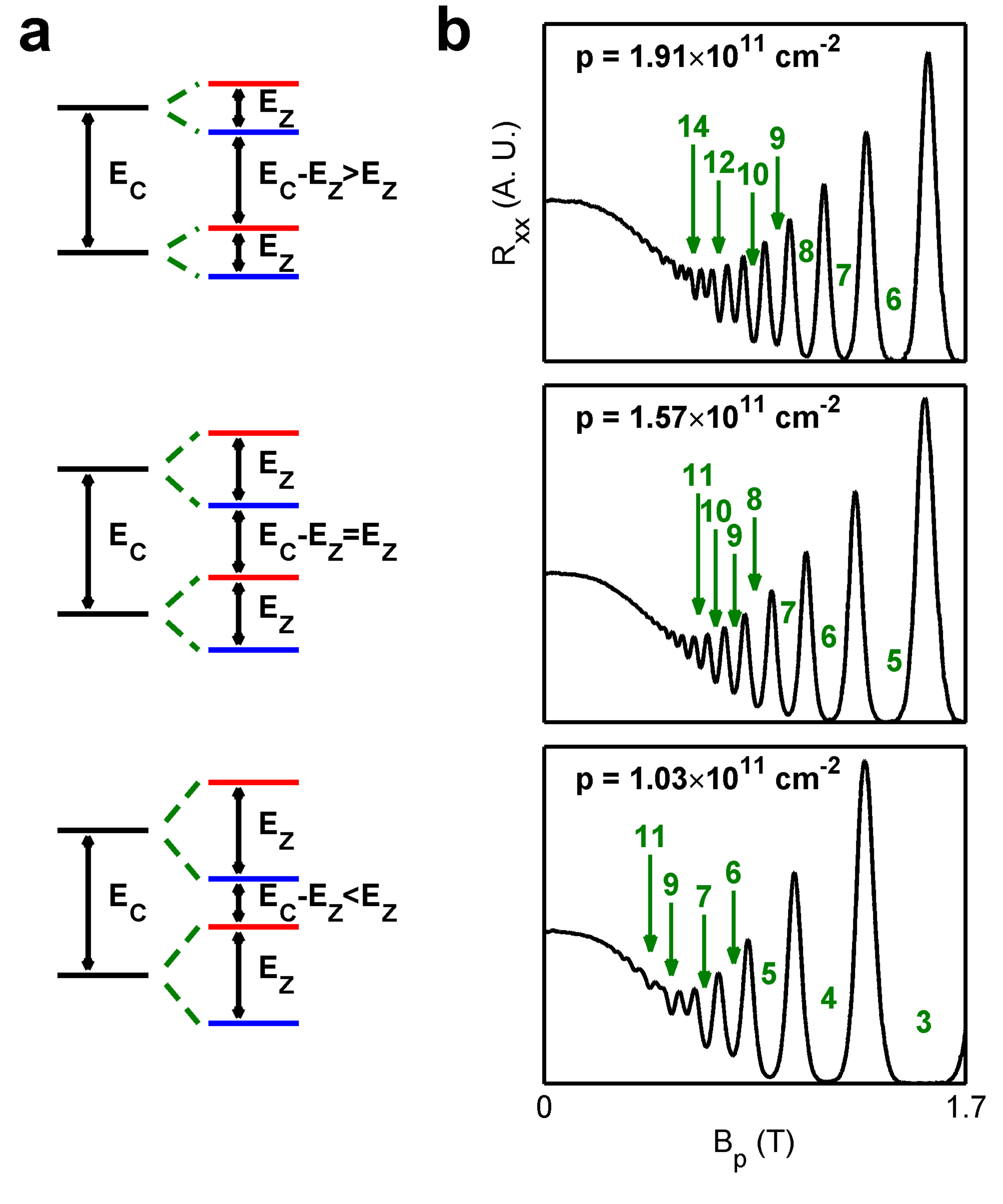}}
\caption{\label{Fig1} (\textbf{a}) The Landau level ladders of a 2D system in a perpendicular magnetic field at $E_{z}/E_{c}<0.5$, $E_{z}/E_{c}\sim0.5$, $E_{z}/E_{c}>0.5$.  Here, $E_{z}$ is the Zeeman splitting and $E_{c}$ is the cyclotron gap.  The red (blue) lines are spin up (down) Landau levels.  (\textbf{b}) $R_{xx}$ of 2D holes in Ge at $T=$ 0.3 K.  From top to bottom, the densities are: $1.91\times 10^{11}$ cm$^{-2}$, which shows strong even states, $1.57\times 10^{11}$ cm$^{-2}$, which shows equally strong even and odd states, and $1.03\times 10^{11}$ cm$^{-2}$, which shows strong odd states.  }
\end{figure}

\begin{figure}[h]
\resizebox{6 in}{!}{\includegraphics{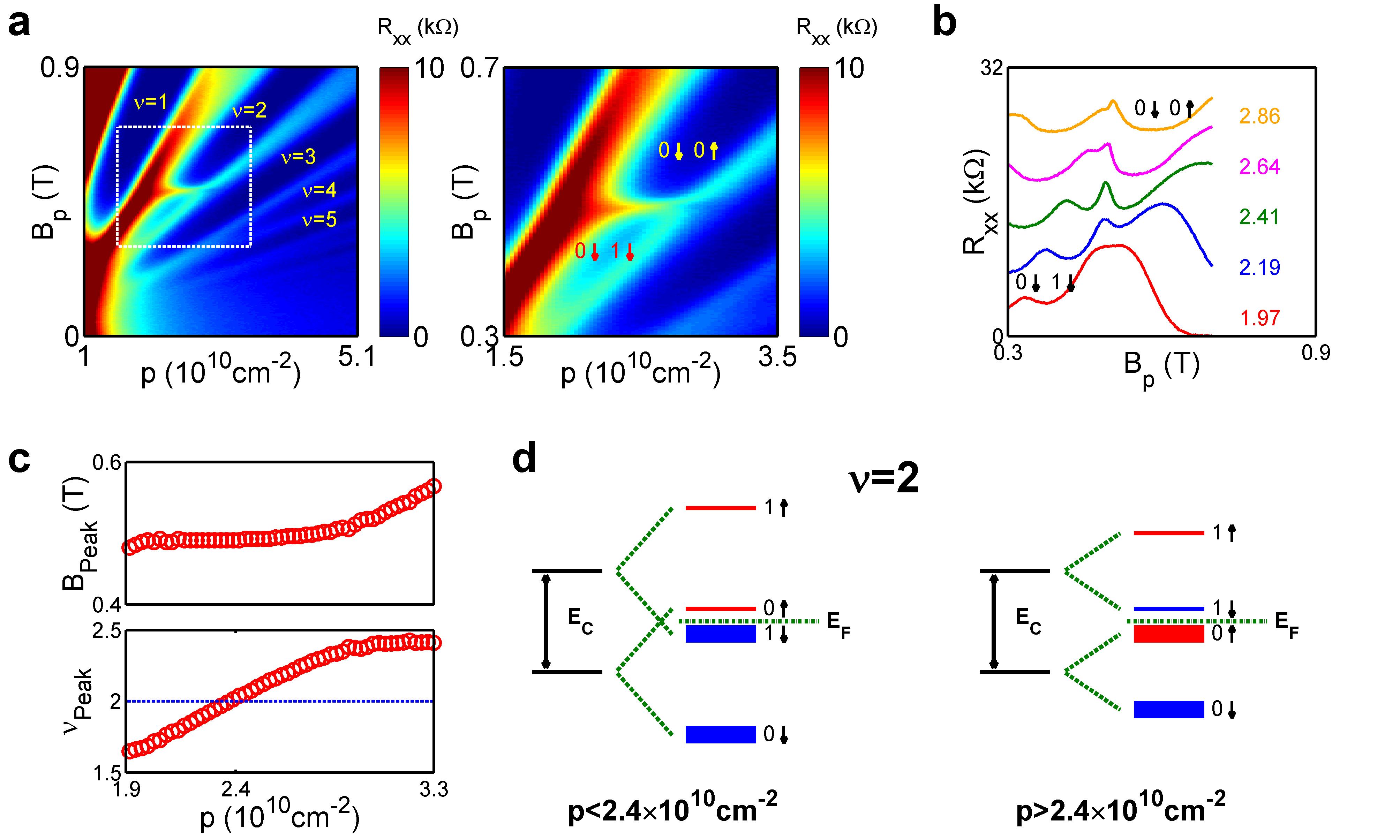}}
\caption{\label{Fig2} (\textbf{a}) $R_{xx}$ of 2D holes in Ge as a function of $p$ and $B_{p}$.  Prominent filling factors are labeled in yellow.  A zoom-in view of the quantum Hall ferromagnetic transition is shown on the right panel.  The lowest two filled Landau levels before and after the transition are labeled.  (\textbf{b}) Line cuts along the $B_{p}$ direction at five densities near the transition, showing the evolution of the resistance peak due to magnetic domains.  (\textbf{c}) The position and the corresponding filling factor of the resistance peak as a function of $p$. (\textbf{d}) The alignment of the four lowest Landau levels before and after the transition at $p=2.4\times10^{10}$ cm$^{-2}$.  The Fermi level is between the second and the third lowest spin-polarized Landau levels, corresponding to $\nu=2$.}
\end{figure}

\begin{figure}[h]
\resizebox{6 in}{!}{\includegraphics{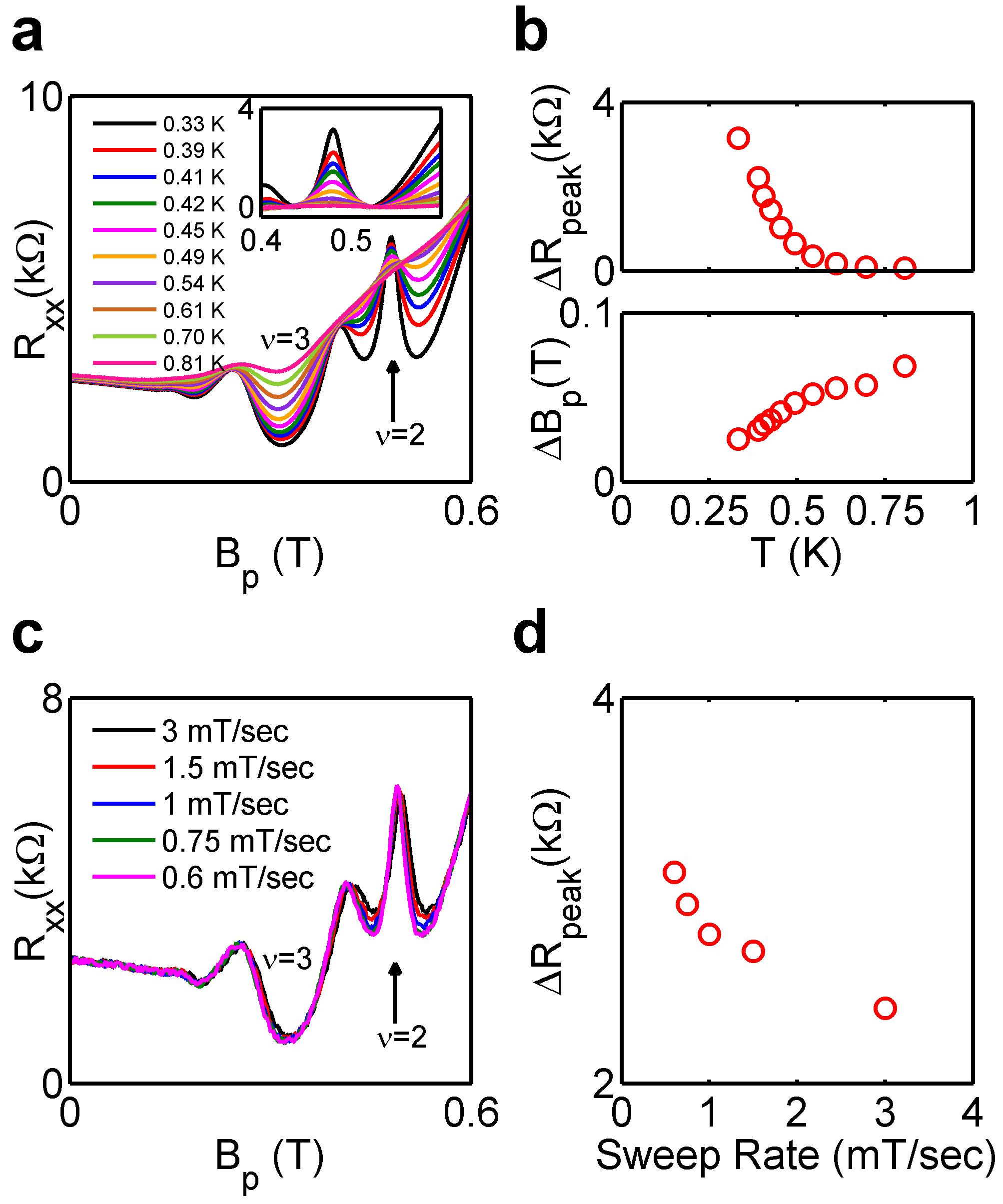}}
\caption{\label{Fig3} (\textbf{a}) Temperature dependence of $R_{xx}$ at $p\sim2.4\times10^{10}$ cm$^{-2}$.  The inset shows the peak with a linear background subtracted from $R_{xx}$ at each temperature.  (\textbf{b}) Amplitude and width of the resistance peak as functions of temperature are shown in the top and bottom pannels, respectively.  Errors are smaller than the size of the symbols.  (\textbf{c}) Magnetic-field sweep rate dependence of $R_{xx}$ at $p\sim2.4\times10^{10}$ cm$^{-2}$. (\textbf{d}) The peak height as a function of the sweep rate.  Errors are smaller than the size of the symbols.}
\end{figure}

\end{document}